\documentclass[bibyear]{aa}  
\usepackage{natbib}
\bibpunct{(}{)}{;}{a}{}{,}

\usepackage{graphicx}
\usepackage[varg]{txfonts}
\usepackage{hyperref}
\hypersetup{
    colorlinks=true,
    linkcolor=blue,
    urlcolor=blue,   
    citecolor=blue,
}

\usepackage{bm}

\begin{document}

\title{Blazars define a stable celestial reference frame}

\author{N.~J.~Secrest\inst{1} \and S.~Lambert\inst{2}}

\institute{Celestial Reference Frame Department, U.S. Naval Observatory, 3450 Massachusetts Ave NW, Washington, DC 20392, USA \\ \email{nathan.j.secrest.civ@us.navy.mil} \and LTE, Observatoire de Paris, Universit\'e PSL, CNRS UMR8255, Sorbonne Universit\'e, Universit\'e de Lille, LNE, 61 avenue de l’Observatoire, 75014 Paris, France \\ \email{sebastien.lambert@obspm.fr}}
\date{}

\abstract
{Recent work has shown that optical-radio position offsets and radio position variability are inversely correlated with the optical photometric variability of active galactic nuclei (AGN). A key prediction of these findings is that a reference frame constructed using highly photometrically variable AGN should be more stable than a frame that does not account for variability and that photometric variability can be used to optimally weight all sources in order to maximize frame stability.}
{To test this prediction by determining the stability of reference frames constructed using sources selected on photometric variability.}
{Using ICRF3 matched to {\it Gaia}~DR3, we employed a bootstrap method to estimate the multi-epoch stability of frames constructed using AGN selected at varying levels of photometric variability. We fit vector spherical harmonics to the coordinate differences between the three ICRF3 frames ($S/X$, $K$, and $X/Ka$) and {\it Gaia}~DR3 and quantified the statistical dispersion as a function of blazar-like (high variability), quasar-like (low variability), and intermediate-variability class.}
{An $S/X$ reference frame constructed using blazars exceeds the stability of a frame constructed with quasars by a factor of 6 and is twice as stable as the ICRF3 defining sources. At $K$ and $X/Ka$, a blazar-based frame matches or exceeds the stability of the defining sources by a factor of $1.4$ in the case of $X/Ka$ and exceeds the stability of a frame based on quasars by over a factor of 2 in both cases. The smaller improvement at $K$ and $X/Ka$ is likely because sources selected at higher frequency are more likely to be blazars. We derived a variability-based astrometric covariance scaling method that results in factor of 2 reduction in frame distortions and instabilities between ICRF3\,$S/X$ and {\it Gaia}~DR3, with a mild improvement for ICRF3\,$K$ but no difference for ICRF3\,$X/Ka$, which is dominated by known distortions.}
{A celestial reference frame defined using blazars is likely to be stable over time. This confirms the prediction that an optimal weighting of the link between the optical and radio celestial reference frames is enabled by accounting for photometric variability.}
{}

\keywords{Reference systems -- quasars: general -- Galaxies: jets -- Galaxies: active -- Astrometry}

\maketitle

\nolinenumbers

\section{Introduction}

At the turn of the 21st century, the fundamental frame of reference used to define astronomical positions and the orientation of the Earth was created using bright, compact, extragalactic radio sources monitored with very long baseline interferometry \citep[VLBI;][]{1995A&A...303..604A}. The geodetic VLBI technique is especially suitable for realizing a celestial reference frame because the positions of radio sources on the sky and the positions of the radio telescopes on Earth are solved for simultaneously as a global solution that directly links the celestial reference frame to the terrestrial reference frame, yielding the Earth orientation parameters. 

Following the first International Celestial Reference Frame \citep[ICRF1;][]{1998AJ....116..516M}, the ICRF has undergone a process of gradual expansion and refinement, initially being defined solely using the geodetic 2.3 and 8.4~GHz bands (respectively $S$/$X$), but later also defined at 24~GHz ($K$) and 32~GHz ($Ka$) with the release of the third and current ICRF \citep[ICRF3;][]{2020A&A...644A.159C}. In comparison to ICRF1 and ICRF2 \citep{2015AJ....150...58F}, ICRF3 has fewer distortions in the frame because the defining sources were selected to be distributed more uniformly across the celestial sphere. Additionally, ICRF3 was corrected for Galactic aberration, with \citet{2020A&A...644A.159C} adopting a value of 5.8~$\mu$as~yr$^{-1}$ \citep[see also][]{MacMillanFeyGipson2019}. The rotational stability of the ICRF is typically estimated to be a few tens of $\mu$as, with 30~$\mu$as estimated for ICRF3\,$S/X$ \citep{2020A&A...644A.159C}.

This error, however, is not due to measurement noise from a finite number of defining sources. As we emphasize here, most of the frame instability is systematic and likely to be mainly caused by effects inherent to the astrophysical properties of the defining sources. One such effect that is commonly discussed in the literature is the frequency-dependent core shift \citep{2008A&A...483..759K}, which causes astrometric offsets of $\sim$~0.1--1~mas on average in phase-delay astrometric measurements, and offsets in group delay-based positions such as those that were used to construct the ICRF when the jet is not in equipartition \citep[see, e.g., the discussion in][]{2009A&A...505L...1P}. The majority of sources that have extended structure at VLBI scales show evidence for core shift variability \citep{2019MNRAS.485.1822P}, and the deviations from the core distance-frequency dependence $r(\nu)~\propto \nu^{-1}$ equipartition prediction \citep{1979ApJ...232...34B} regularly exceed $\sim0.3$~mas \cite[see \citealt{2011A&A...532A..38S} and corresponding discussion in][]{2019MNRAS.485.1822P}. As noted by \citet{2025ApJS..276...38P}, particle flares in the jet, which cause core shift variability, disrupt equipartition. As these flares are a universal feature of radio jets, group delays may not in general be considered impervious to core shift. \citet{2024ApJS..274...28C} recently showed based on $S/X$ position time series that ICRF3 sources exhibit a typical excess astrometric noise of $\sim$~0.1--1~mas (their Fig.~8), and half of the defining sources showed excess noise greater than $\sim0.2$~mas (their Table~1).

The selection of defining sources for a fundamental reference frame such as the ICRF therefore requires care to select the most stable and well-behaved objects. In ICRF1 and ICRF2, this was done by selecting sources with observing sessions that covered at least 2~yr, with cuts on formal VLBI position uncertainties made for ICRF1 and position stability and source structure made for ICRF2. In ICRF3, special emphasis was placed on selecting defining sources based on uniformity of sky distribution. Position stability over several epochs, however, is no guarantee that a source will remain astrometrically well behaved. An extraordinary example of this is 3C\,48, which, despite a sub-milliarcsecond astrometric position uncertainty in ICRF3 for 18 observing sessions preceding 2017~September~7, exhibited a shift of 57~mas in declination shortly thereafter.\footnote{The astrometric time series for 3C\,48 and other VLBI sources, presented by \citet{2024ApJS..274...28C}, may be found at \url{https://crf.usno.navy.mil/quarterly-vlbi-solution}.} Astrometric shifts of similar magnitude are known for at least three other sources \citep{2022MNRAS.512..874T}, which raises the disturbing prospect that an apparent astrometric stability is of limited benefit when choosing sources for a reference frame. None of the four sources studied by \citet{2022MNRAS.512..874T} is defining, but \citet{2024ApJS..274...28C} recently identified several ICRF3 defining sources with excess astrometric dispersions of $\sim1$~mas (see their Table~2).

It is clear that in order to establish a long-term stable ICRF, it is desirable to determine not just which sources, but which type of sources are astrometrically better behaved across epochs. This is especially important now that the ICRF is multifrequency, with {\it Gaia}-CRF3 \citep{2022A&A...667A.148G} being the official instantiation of the ICRS system at visual (optical) wavelengths.\footnote{\url{https://drive.google.com/file/d/1a-gZ8NZQtGpKart-UOSa3WgnxVp4ls_3/view?pli=1}} Frequency-dependent source physics, such as the aforementioned core shift, limit efforts to create a consistent multiwavelength reference frame.\footnote{\url{https://iau.org/WG329/WG329/Home.aspx}} Understanding the processes that create frequency-dependent positions, such as synchrotron opacity, contamination by the host galaxy, or offsets between the optically bright accretion disk and the radio core should therefore be at the forefront of reference frame research.

Toward this goal, \citet{2022ApJ...939L..32S} showed that selection on photometric variability has a profound effect on the prevalence of offsets between the {\it Gaia} optical and ICRF radio positions, which affect about $10\%-15\%$ of the ICRF3 sources overall. While about $20\%-25\%$ of the least photometrically variable ICRF sources exhibit optical-radio offsets, this prevalence drops to just a few percent in objects with a photometric variability exceeding $\sim20\%$ \citep[see Fig.~2 in][]{2022ApJ...939L..32S}. Objects with large fractional photometric variabilities are likely blazars: sources whose relativistic jet is pointed close to our line of sight. Through a comparison of optical color and $\gamma$-ray counterpart incidence, \citet{2022ApJ...939L..32S} showed that this is unambiguously the case for ICRF objects. This result was further strengthened by an analysis of geodetic VLBI radio position variability published by \citet{2024A&A...684A..93L}, who showed that VLBI position stability is indeed correlated with photometric variability, with blazars such as flat-spectrum radio quasars (FSRQs) and BL~Lacs having the most stable positions \citep[see also][]{2025A&A...695A.135L}.

These results are naturally interpreted as arising from two factors inherent to blazars. First, the direction of the jet near to the line of sight minimizes projected position offsets, so the apparent positions of blazars are largely unaffected by the structural variations of their jets. Second, the small viewing angle strongly boosts the apparent luminosity of the jet at all frequencies, so the contributions of nonjet structures such as the accretion disk are proportionally smaller, which further improves the positional consistency. A key prediction made by \citet{2022ApJ...939L..32S} is that, as long as the direction of the jet remains close to the line of sight, blazars are positionally stable, which makes them ideal objects for a stable reference frame in the long term. We discuss the merits of the assumption of long-term blazar stability in Sect.~\ref{subsec: longterm} and show that it is likely justified for the purpose of creating a stable reference frame.

In this work, we quantify for the first time the stability of the celestial reference frame as a function of astrophysical source type and show that selection on blazars dramatically improves the stability of the frame. In Sect.~\ref{sec: methods} we detail our formalism and methods. Our results are presented in Sect.~\ref{sec: results}, followed by a discussion in Sect.~\ref{sec: discussion}. Our main conclusions are listed in Sect.~\ref{sec: conclusions}.

\section{Methods} \label{sec: methods}

\subsection{Formalism}
We adopted the $l\leq2$ vector spherical harmonic (VSH) model to parameterize the spin ($l=0$), dipolar glide deformations ($l=1$), and $l=2$ quadrupolar distortions of the reference frame relative to a fiducial standard \citep[see][]{2012A&A...547A..59M}. The VSH equations listed by \citet[][see also \citealt{2020A&A...644A.159C}]{2013A&A...559A..95T} were solved using generalized least squares,

\begin{equation} \label{eq: gls}
\hat{\vec{\varepsilon}} = \left(\vec{A}^T \vec{\Sigma}^{-1} \vec{A} \right)^{-1} \vec{A}^T \vec{\Sigma}^{-1} \vec{\Delta},
\end{equation}

\noindent where $\vec{A}$ is a $2n \times 16$ array of design matrices for $n$ sources, $\vec{\Sigma}$ is the $2n \times 2n$ block diagonal matrix of position covariances (the covariance terms between different sources were set to zero), and $\vec{\Delta}$ is a $2n\times1$ column matrix of the position offsets in right~ascension and declination. The formal covariance of Eq.~\ref{eq: gls} is

\begin{equation} \label{eq: gls_cov_formal}
\Sigma_{\hat{\vec{\varepsilon}}} = \left(\vec{A}^T \vec{\Sigma}^{-1} \vec{A} \right)^{-1}.
\end{equation}

\noindent Equation~\ref{eq: gls_cov_formal} assumes that the data covariances are both accurate and Gaussian. This is not usually true of VLBI data, for which the practice has been to apply a scale factor of 1.5 to the formal errors, followed by the addition in quadrature of some noise floor \citep[in the case of ICRF3\,$S/X$, 30~$\mu$as;][]{2020A&A...644A.159C}. The noise floor was derived using decimation tests in which the data were split by epoch to estimate positional variability. For the purposes of this work, however, these post hoc uniformly applied corrections serve only to dilute the astrometric quality differences between source types, as we show below.

\subsection{The bootstrap} \label{subsec: bootstrap}
In order to quantify the stability of the celestial reference frame as a function of astrophysical source type, we used random sampling with replacement \citep[the bootstrap;][]{efron79}. The bootstrap is a means of estimating the sampling distribution of an estimator given a particular set of data, so it therefore allows estimation of the frame stability performance of a source class as a population. Bootstrapping also has another advantage in that it provides a means of estimating the effect of source variability within a class. This is because individual sources are uncorrelated, so the state of one source at the time of a particular observation may be taken as representative of the state of a different source at the time of some future observation that occurs beyond the timescale on which the observed position of a source is correlated with its previously observed position. Timeseries analysis of the VLBI positions of ICRF objects suggests that this timescale is months to years \citep{2024ApJS..274...28C}. Given that there are physical processes in AGN that cause apparent position variation with time and wavelength, such as variations in synchrotron self-absorption, variations in Doppler factor, flares in the shock structure of the jet, superluminal motion, or accretion disk variability, and given that these physical processes are uncorrelated between sources, so different sources are in different states at different times (e.g.\ a flare in one source is unrelated to a flare in another source), it follows that comparison of different but randomly selected sets of sources is logically equivalent to comparison of one set of sources at a different epoch. The variation in frame orientation between samples beyond the dispersion expected from measurement or statistical error is directly due to source positional variability. This argument suggests, however, that frame stability is properly tested through random sampling without replacement. Unfortunately, there are not enough sources to create the subsamples for this test. For example, there are 180 blazar-like objects in the sample of 1938 ICRF3-{\it Gaia} sources developed by \citet{2022ApJ...939L..32S}, so the subsamples created by random sampling without replacement just 10 times would be dominated by statistical error.

\begin{figure}
\includegraphics[width=\columnwidth]{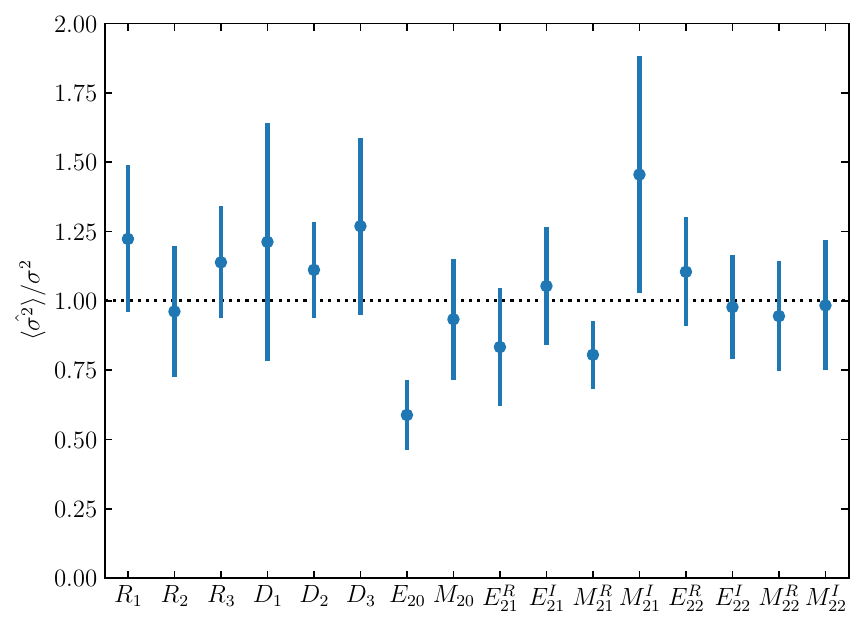}
\caption{Ratio of the average bootstrap sample variance to the true population variance of the 16 terms in the $l\leq2$ VSH model we employed for simulated multi-epoch data in which position offsets vary with epoch. The position offsets were sampled from the empirical distribution of statistically significant ICRF3-{\it Gaia} offsets observed in the sample used by \citet{2022ApJ...939L..32S} and this work. The error bars denote the standard errors of the means.}
\label{fig: bootstrap_demonstration}
\end{figure}

To assess the validity of allowing replacement, we drew 180 random sources from the objects studied by \citet{2022ApJ...939L..32S}. We generated a large number of simulated epochs and assigned to each epoch a randomly oriented offset drawn from the empirical distribution of statistically significant offsets \citep[having covariance-normalized values greater than 5; see][]{2022ApJ...939L..32S} observed between the ICRF3\,$S/X$ and {\it Gaia}~DR3 positions of our sample (Sect.~\ref{subsec: sample}). To reflect the prevalence of significant optical-radio offsets seen in real data, we assigned offsets to 20\% of the sources, which varied randomly between epochs. We then permuted the positions of the simulated offsets by their formal measurement covariances from {\it Gaia} and ICRF3. For a random subset of epochs, we made bootstrap resamples of the sources and fit each resample with the $l\leq2$ VSH model discussed in the previous section. In Fig.~\ref{fig: bootstrap_demonstration}, we compare the variance of the VSH components between bootstrap resamples to the true variance across all epochs. The bootstrap resampling of a set of sources at a particular epoch gives an estimate of the multi-epoch stability of the rotation, glide, and quadrupole parameters that is close to truth on average. It therefore provides a means of assessing the frame stability.

\subsection{Sample} \label{subsec: sample}
The 1938 ICRF3-{\it Gaia} matches used by \citet{2022ApJ...939L..32S} were selected from an initial match between ICRF3\,$S/X$ and {\it Gaia}-CRF3 \citep{2022A&A...667A.148G} to within 100~mas (3142 objects), required to have measured {\it Gaia} $G$-band fractional photometric variability (\texttt{fractional\_variability\_g}, hereafter $F_\mathrm{var}$; 2426 objects), no evidence of source extent or evidence for close secondary sources (1953 objects), and no evidence for spurious parallaxes or proper motions (1938 objects), which can be further evidence for multisources \citep{2022A&A...660A..16S, 2022ApJ...933...28M}. As stated previously, 180 have $F_\mathrm{var} > 0.4$, which \citet{2022ApJ...939L..32S} showed are likely pure line-of-sight blazars, 1455 have $F_\mathrm{var} < 0.2$ and have properties consistent with radio-quiet quasars, so they likely do not have jets pointed as close to the line of sight, and 303 have $0.2 < F_\mathrm{var} < 0.4$, which are likely objects intermediate between quasars and pure blazars in which the jet is significantly beamed, but not pointed directly at the observer.

These three fractional variability classes of sources, which we refer to as blazars, quasars, and intermediate sources, formed the basis of the three samples we used. In order to facilitate a meaningful comparison of the frame stability as a function of source class, however, we required that all three samples have the same size to control for purely statistical error and to be as closely matched on the sky as possible to control for the susceptibility to particular distortions. The high $F_\mathrm{var}$ blazar-like category, which has the fewest objects, set the sample size $n$, and we therefore selected the $n$ closest quasar-like and intermediate sources to the blazars. For sources with $S/X$, $K$, and $X/Ka$ positions, $n=180$, $n=75$, and $n=68$, respectively. To facilitate consideration of a reference frame with defining sources selected from bona fide variability-selected blazars, we also produced a matched sample of the ICRF3 defining sources.

Position offsets were taken between {\it Gaia} and the ICRF3 source positions. At the epoch of {\it Gaia}~DR3 (2016.0), {\it Gaia} positions of unresolved sources generally have reliable position covariances, with uncertainties likely underestimated by $\sim6\%$ \citep[see discussion in][]{2022ApJ...939L..32S}. Since the ICRF3 and {\it Gaia} data we used differ in epoch by only 0.5~yr, the effect of Galactic aberration, which creates a glide of about 5~$\mu$as~yr$^{-1}$, was not considered. Because the post~hoc corrections to the VLBI position uncertainties applied to ICRF3 \citep[][Table~6]{2020A&A...644A.159C} only dilute the observable (the presence of systematic position error due to astrophysical effects), we restored the original VLBI position uncertainties.

Finally, we note that while high fractional variability reliably selects for bona fide blazars, this metric is probably substantially incomplete because many genuine blazars did not undergo flaring during the epochs in which {\it Gaia} observed them. Consequently, many genuine blazars undoubtedly have $F_\mathrm{var}$ values below the cut employed by \citet{2022ApJ...939L..32S} and \citet{2024A&A...684A..93L}. Employing other criteria such as cuts made on the \texttt{qso\_variability} parameter \citep{2011AJ....141...93B} provided in {\it Gaia}~DR3, inclusion of additional data such as multi-epoch photometry from surveys such as the Zwicky~Transient~Facility \citep{2019PASP..131a8002B} or the Near-Earth Object WISE (NEOWISE) mission \citep{2011ApJ...731...53M,2014ApJ...792...30M}, or consideration of $\gamma$-ray detection by the \textit{Fermi} Large~Area~Telescope \citep{2022ApJS..260...53A} as a blazar indicator will improve the completeness significantly. The size of the blazar sample we explored should therefore not be construed as the limit set by nature.

\section{Results} \label{sec: results}
\subsection{Effect of photometric variability class}
\begin{figure*}
\includegraphics[width=\textwidth]{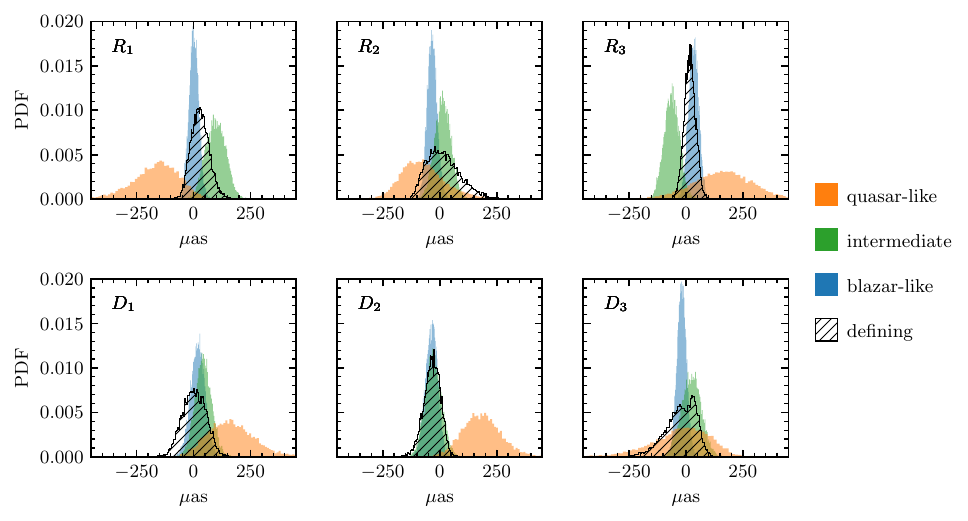}
\caption{Distribution of $l=0$ (rotation) and $l=1$ (glide) values between ICRF3\,$S/X$ and {\it Gaia}~DR3, separating objects by fractional variability into quasar-like ($F_\mathrm{var} < 0.2$), intermediate ($0.2 < F_\mathrm{var} < 0.4$), and blazar-like ($F_\mathrm{var} > 0.4$) samples. A dramatic improvement in frame stability and rigidity is enabled by selection on blazars, exceeding even the sources used to define the ICRF3 frame.}
\label{fig: rdem}
\end{figure*}

\begin{figure*}
\includegraphics[width=\textwidth]{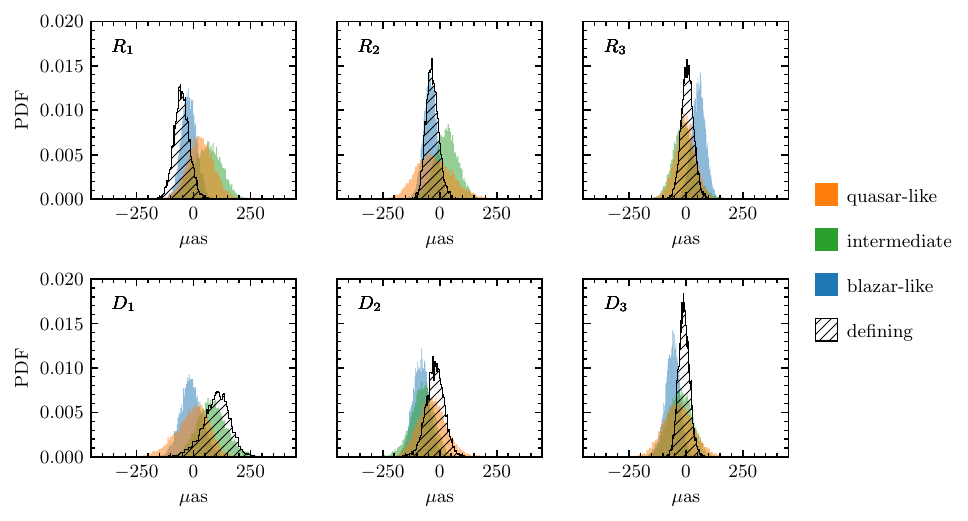}
\caption{Distribution of $l=0$ (rotation) and $l=1$ (glide) values between ICRF3\,$K$ and {\it Gaia}~DR3, separating objects by fractional variability into quasar-like ($F_\mathrm{var} < 0.2$), intermediate ($0.2 < F_\mathrm{var} < 0.4$), and blazar-like ($F_\mathrm{var} > 0.4$) samples.}
\label{fig: rdem k}
\end{figure*}

\begin{figure*}
\includegraphics[width=\textwidth]{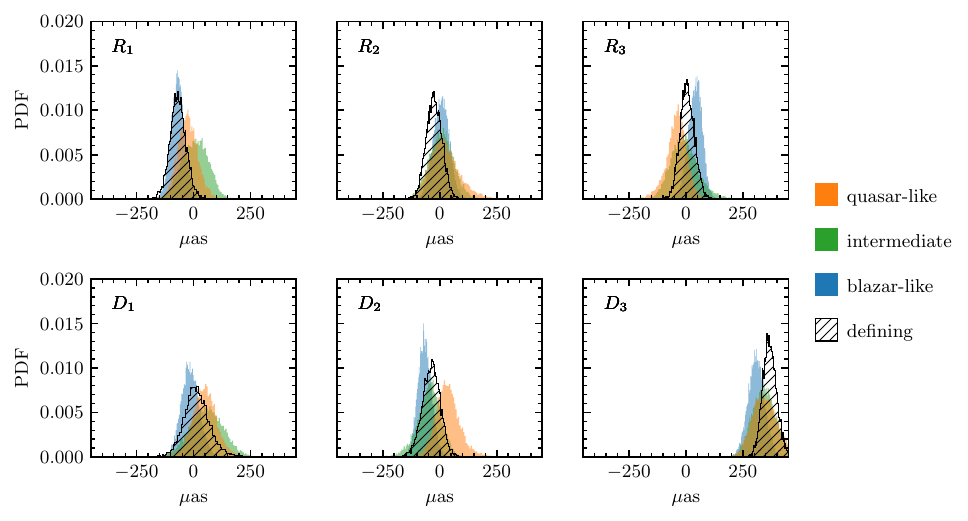}
\caption{Distribution of $l=0$ (rotation) and $l=1$ (glide) values between ICRF3\,$X/Ka$ and {\it Gaia}~DR3, separating objects by fractional variability into quasar-like ($F_\mathrm{var} < 0.2$), intermediate ($0.2 < F_\mathrm{var} < 0.4$), and blazar-like ($F_\mathrm{var} > 0.4$) samples. The large value of the $D_3$ term is a known issue and was discussed by \citet{2020A&A...644A.159C}.}
\label{fig: rdem xka}
\end{figure*}

The results of the bootstrap are shown in Figs.~\ref{fig: rdem}, \ref{fig: rdem k}, and \ref{fig: rdem xka} for $S/X$, $K$, and $X/Ka$, respectively. The difference in the frame stability and rigidity between frames defined using low-variability quasar-like sources and high-variability blazar-like sources for the $S/X$ band is striking, with the difference being less but still notable for $K$ and $X/Ka$. The mean offsets of the bootstrap samples from zero, most notable at $S/X$, suggest that selection on quasar-like objects also introduces significant frame distortions. The intrinsic dispersions of the bootstrap samples of all 16 VSH parameters are given in Tables~\ref{tab: rdem sx}, \ref{tab: rdem k}, and \ref{tab: rdem xka} for $S/X$, $K$, and $X/Ka$. The intrinsic dispersion was estimated by subtracting the statistical error from the total observed dispersion in quadrature. We estimated the statistical error numerically by permuting a fiducial position of each source by the formal covariances of the VLBI and {\it Gaia} positions to create a simulated version of the source, where the observed VLBI-{\it Gaia} offset is entirely attributable to observation error. We used the {\it Gaia} positions as the fiducial positions for convenience because they remained the same for the $S/X$, $K$, and $X/Ka$ analyses. Between $S/X$ and {\it Gaia}, blazars define a frame that is about six times more stable than quasar-like objects on average, and they are over twice as stable as the defining sources themselves. For $K$, blazars are similarly as stable as the defining sources, but twice as stable as quasars. At $X/Ka$, blazars again outperform the defining sources and are again twice as stable as the quasars. A reference frame defined using blazars is therefore consistently and significantly more stable than a frame defined using other object types, and it can be expected to significantly improve upon the current set of defining sources for $S/X$ and $X/Ka$.

\begin{table}
\caption{Intrinsic instabilities in rotations, glides, and quadrupole distortions between $S/X$ and {\it Gaia}.} \label{tab: rdem sx}
\begin{tabular}{crrrr}
\hline
\hline \\ [-0.36cm]
Parameter & $\sigma_\textrm{quasar-like}$ & $\sigma_\textrm{intermediate}$ & $\sigma_\textrm{blazar-like}$ & $\sigma_\textrm{defining}$ \\ [0.05cm]
 \hline \\[-0.33cm]
$R_1$ & 108 & 40 & 19 & 38 \\ [0.05cm]
$R_2$ & 102 & 32 & 18 & 74 \\ [0.05cm]
$R_3$ & 131 & 29 & 19 & 24 \\ [0.05cm]
$D_1$ & 103 & 29 & 26 & 49 \\ [0.05cm]
$D_2$ & 85 & 33 & 24 & 35 \\ [0.05cm]
$D_3$ & 133 & 40 & 18 & 67 \\ [0.05cm]
$E_{20}$ & 92 & 43 & 20 & 77 \\ [0.05cm]
$M_{20}$ & 105 & 32 & 22 & 32 \\ [0.05cm]
$E^\mathrm{Re}_{21}$ & 117 & 43 & 27 & 82 \\ [0.05cm]
$E^\mathrm{Im}_{21}$ & 182 & 45 & 24 & 45 \\ [0.05cm]
$M^\mathrm{Re}_{21}$ & 174 & 39 & 26 & 51 \\ [0.05cm]
$M^\mathrm{Im}_{21}$ & 191 & 45 & 33 & 60 \\ [0.05cm]
$E^\mathrm{Re}_{22}$ & 90 & 21 & 12 & 24 \\ [0.05cm]
$E^\mathrm{Im}_{22}$ & 72 & 16 & 11 & 17 \\ [0.05cm]
$M^\mathrm{Re}_{22}$ & 85 & 24 & 13 & 26 \\ [0.05cm]
$M^\mathrm{Im}_{22}$ & 77 & 21 & 13 & 48 \\ [0.05cm]
\hline \\ [-0.33cm]
Average & 115 & 33 & 20 & 47 \\
\hline
\hline
\end{tabular}
\tablefoot{Values given in $\mu$as, estimated as $\sigma^2 = \sigma_\mathrm{total}^2 - \sigma_\mathrm{stat}^2$, where $\sigma_\mathrm{total}$ is the dispersion observed in the samples, and $\sigma_\mathrm{stat}$ is the dispersion due to sample size.}
\end{table}

\begin{table}
\caption{Intrinsic instabilities in rotations, glides, and quadrupole distortions between $K$ and {\it Gaia}.}\label{tab: rdem k}
\begin{tabular}{crrrr}
\hline
\hline \\ [-0.36cm]
Parameter & $\sigma_\textrm{quasar-like}$ & $\sigma_\textrm{intermediate}$ & $\sigma_\textrm{blazar-like}$ & $\sigma_\textrm{defining}$ \\ [0.05cm]
 \hline \\[-0.33cm]
$R_1$ & 55 & 59 & 28 & 26 \\ [0.05cm]
$R_2$ & 78 & 47 & 23 & 22 \\ [0.05cm]
$R_3$ & 42 & 46 & 24 & 22 \\ [0.05cm]
$D_1$ & 69 & 62 & 36 & 37 \\ [0.05cm]
$D_2$ & 59 & 46 & 30 & 32 \\ [0.05cm]
$D_3$ & 63 & 46 & 27 & 16 \\ [0.05cm]
$E_{20}$ & 90 & 51 & 26 & 32 \\ [0.05cm]
$M_{20}$ & 50 & 51 & 27 & 34 \\ [0.05cm]
$E^\mathrm{Re}_{21}$ & 95 & 48 & 37 & 29 \\ [0.05cm]
$E^\mathrm{Im}_{21}$ & 73 & 57 & 33 & 26 \\ [0.05cm]
$M^\mathrm{Re}_{21}$ & 82 & 61 & 33 & 38 \\ [0.05cm]
$M^\mathrm{Im}_{21}$ & 86 & 78 & 45 & 48 \\ [0.05cm]
$E^\mathrm{Re}_{22}$ & 26 & 28 & 13 & 15 \\ [0.05cm]
$E^\mathrm{Im}_{22}$ & 35 & 25 & 15 & 15 \\ [0.05cm]
$M^\mathrm{Re}_{22}$ & 42 & 30 & 19 & 10 \\ [0.05cm]
$M^\mathrm{Im}_{22}$ & 39 & 25 & 18 & 12 \\ [0.05cm]
\hline \\ [-0.33cm]
Average & 61 & 48 & 27 & 26 \\
\hline
\hline
\end{tabular}
\tablefoot{Values given in $\mu$as, estimated as $\sigma^2 = \sigma_\mathrm{total}^2 - \sigma_\mathrm{stat}^2$, where $\sigma_\mathrm{total}$ is the dispersion observed in the samples, and $\sigma_\mathrm{stat}$ is the dispersion due to sample size.}
\end{table}

\begin{table}
\caption{Intrinsic instabilities in rotations, glides, and quadrupole distortions between $X/Ka$ and {\it Gaia}.}\label{tab: rdem xka}
\begin{tabular}{crrrr}
\hline
\hline \\ [-0.36cm]
Parameter & $\sigma_\textrm{quasar-like}$ & $\sigma_\textrm{intermediate}$ & $\sigma_\textrm{blazar-like}$ & $\sigma_\textrm{defining}$ \\ [0.05cm]
 \hline \\[-0.33cm]
$R_1$ & 37 & 51 & 19 & 28 \\ [0.05cm]
$R_2$ & 60 & 48 & 28 & 26 \\ [0.05cm]
$R_3$ & 42 & 58 & 23 & 26 \\ [0.05cm]
$D_1$ & 44 & 67 & 0 & 39 \\ [0.05cm]
$D_2$ & 47 & 48 & 18 & 31 \\ [0.05cm]
$D_3$ & 54 & 45 & 30 & 25 \\ [0.05cm]
$E_{20}$ & 62 & 56 & 29 & 40 \\ [0.05cm]
$M_{20}$ & 42 & 57 & 13 & 36 \\ [0.05cm]
$E^\mathrm{Re}_{21}$ & 62 & 52 & 37 & 27 \\ [0.05cm]
$E^\mathrm{Im}_{21}$ & 54 & 52 & 22 & 30 \\ [0.05cm]
$M^\mathrm{Re}_{21}$ & 74 & 64 & 34 & 49 \\ [0.05cm]
$M^\mathrm{Im}_{21}$ & 55 & 78 & 13 & 57 \\ [0.05cm]
$E^\mathrm{Re}_{22}$ & 24 & 36 & 15 & 20 \\ [0.05cm]
$E^\mathrm{Im}_{22}$ & 35 & 33 & 13 & 16 \\ [0.05cm]
$M^\mathrm{Re}_{22}$ & 31 & 32 & 16 & 14 \\ [0.05cm]
$M^\mathrm{Im}_{22}$ & 38 & 24 & 20 & 17 \\ [0.05cm]
\hline \\ [-0.33cm]
Average & 48 & 50 & 21 & 30 \\
\hline
\hline
\end{tabular}
\tablefoot{Values given in $\mu$as, estimated as $\sigma^2 = \sigma_\mathrm{total}^2 - \sigma_\mathrm{stat}^2$, where $\sigma_\mathrm{total}$ is the dispersion observed in the samples, and $\sigma_\mathrm{stat}$ is the dispersion due to sample size.}
\end{table}

\subsection{Optimal weighting based on variability} \label{subsec: weighting}
\begin{figure}
\includegraphics[width=\columnwidth]{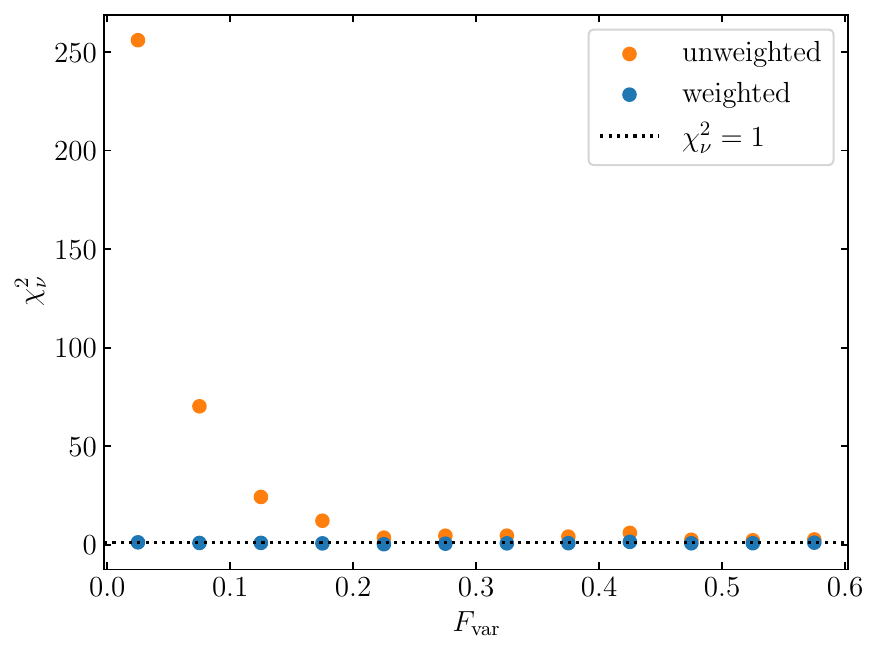}
\caption{Reduced chi-squared statistic of the 16-parameter VSH model as a function of fractional variability using the ICRF3\,$S/X$ and {\it Gaia}~DR3 positions, showing that the variability-based covariance scaling given by Eq.~\ref{eq: covariance weighting} correctly weights the source positions.}
\label{fig: rchi2_vs_fvar_sx}
\end{figure}

\begin{figure}
\includegraphics[width=\columnwidth]{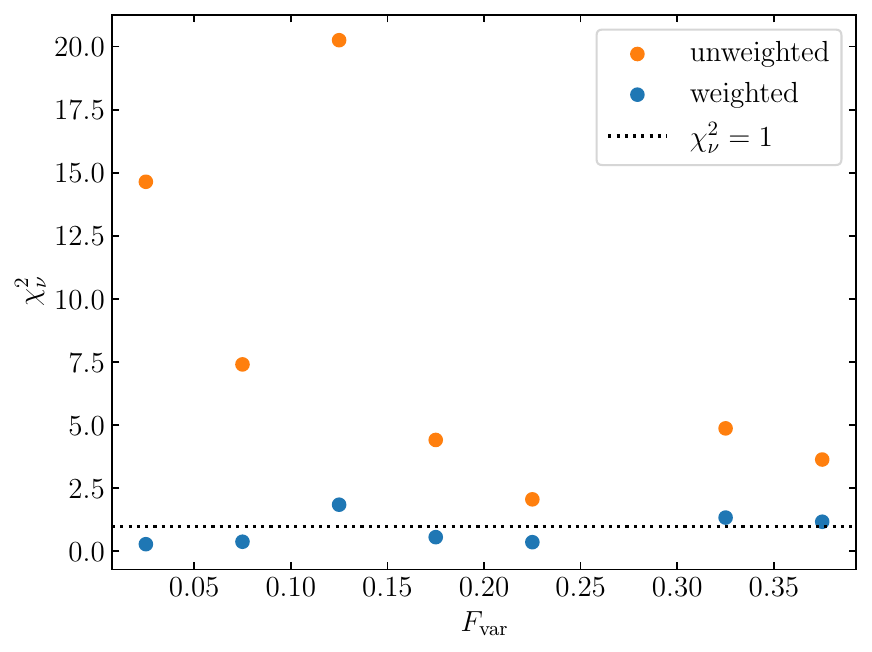}
\caption{Reduced chi-squared statistic of the 16-parameter VSH model as a function of fractional variability using the ICRF3\,$K$ and {\it Gaia}~DR3 positions, showing that the variability-based covariance scaling given by Eq.~\ref{eq: covariance weighting} correctly weights the source positions.}
\label{fig: rchi2_vs_fvar_k}
\end{figure}

\begin{figure}
\includegraphics[width=\columnwidth]{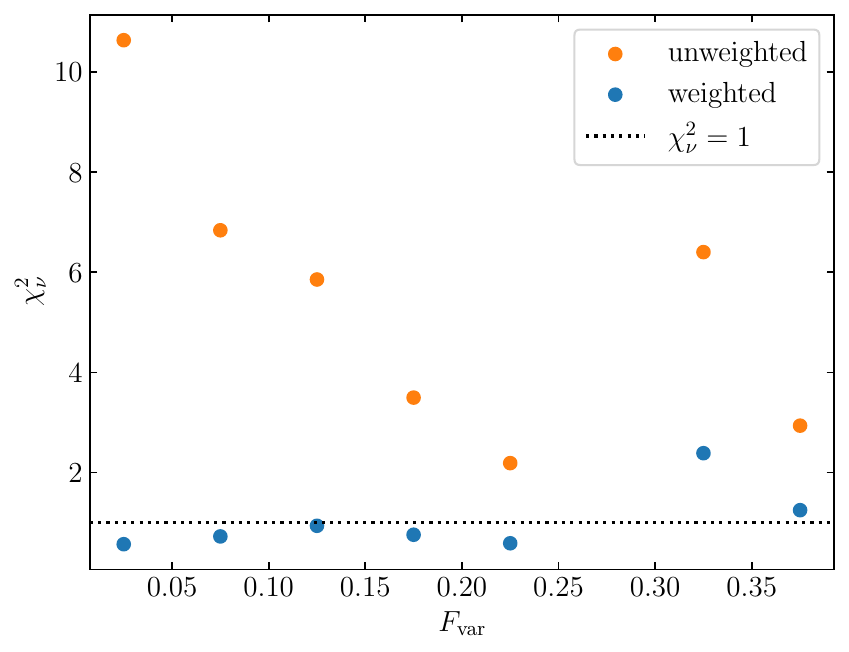}
\caption{Reduced chi-squared statistic of the 16-parameter VSH model as a function of fractional variability using the ICRF3\,$X/Ka$ and {\it Gaia}~DR3 positions, showing that the variability-based covariance scaling given by Eq.~\ref{eq: covariance weighting} correctly weights the source positions.}
\label{fig: rchi2_vs_fvar_xka}
\end{figure}

A key prediction of \citet{2022ApJ...939L..32S} was that variability can be used to optimally weight the optical-radio celestial reference frame link. To test this prediction, we binned the full sample of ICRF3-{\it Gaia} matches from \citet{2022ApJ...939L..32S} by $F_\mathrm{var}$ and fit the VSH model to each bin, returning the reduced chi-squared fit statistic as a function of $F_\mathrm{var}$. As shown in Figs.~\ref{fig: rchi2_vs_fvar_sx}, \ref{fig: rchi2_vs_fvar_k}, and \ref{fig: rchi2_vs_fvar_xka}, the fit statistic shows a clear dependence on $F_\mathrm{var}$, dramatically improving for sources in the intermediate or blazar-like classes ($F_\mathrm{var} > 0.2$). This suggests a means of weighting the covariance matrix in Eq.~\ref{eq: gls} and Eq.~\ref{eq: gls_cov_formal}. 

We found that a remarkably simple scaling can be applied to the covariance matrix $\vec{\Sigma}$. Noting that $F_\mathrm{var}$ is dimensionless, we determined a weighting vector $\vec{w} = \vec{F}_\mathrm{var}^{-x}$ such that

\begin{equation} \label{eq: covariance weighting}
\vec{\Sigma} \rightarrow \vec{w}\vec{\Sigma} = \vec{F}_\mathrm{var}^{-x}\vec{\Sigma},
\end{equation}

\noindent where $\vec{F}_\mathrm{var}$ is the vector containing the fractional source variabilities. We chose $x$ such that the survival probability of the $\chi^2$ distribution was close to $\sim0.5$ given the degrees of freedom (d.o.f.) $\nu$. For the original covariances, this probability is zero for all three ICRFs. We find $x_{S/X}=1.61$, $x_{K}=1.15$, and $x_{X/Ka}=0.87$ with d.o.f. $\nu_{S/X}=3860$, $\nu_{K}=846$, and $\nu_{X/Ka}=756$. Because sources selected on $K$ or $Ka$ are a priori more likely to be flat-spectrum blazars, the smaller values of $x$ make sense. 

Using the original covariances, the 16-parameter VSH fit gives formal frame rotational uncertainties smaller than 5\,$\mu$as for $S/X$ and smaller than 10\,$\mu$as for $K$ and $X/Ka$, while after applying Eq.~\ref{eq: covariance weighting}, the formal frame rotational uncertainties reach 13\,$\mu$as for $S/X$, $\sim19$\,$\mu$as for $K$, and $\sim18$\,$\mu$as for $X/Ka$. Thus, weighting the formal astrometric covariance of a source according to its astrophysical properties, namely its photometric variability, yields more realistic frame alignment uncertainties without the need to apply a single post hoc error correction to all sources, as was done by \citet{2020A&A...644A.159C}.

\begin{figure}
\includegraphics[width=\columnwidth]{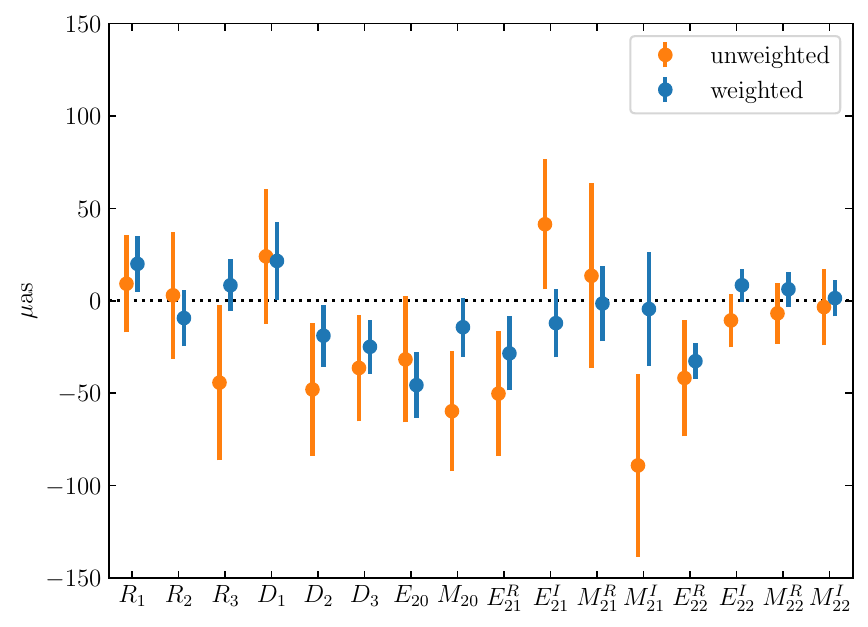}
\caption{Sixteen transformation parameters between ICRF3\,$S/X$ and {\it Gaia}~DR3 before and after weighting by $F_\mathrm{var}$ using Eq.~\ref{eq: covariance weighting} for 1938 sources (d.o.f.~$=3860$). The error bars show the $1\sigma$ dispersion from bootstrapping.}
\label{fig: vsh_offsets_weighting_effect_sx}
\end{figure}

\begin{figure}
\includegraphics[width=\columnwidth]{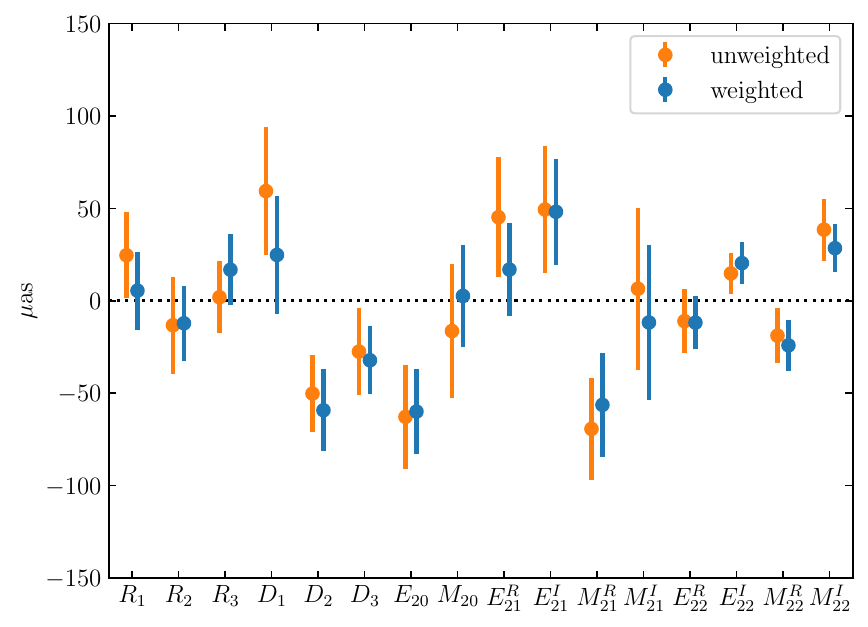}
\caption{Sixteen transformation parameters between ICRF3\,$K$ and {\it Gaia}~DR3 before and after weighting by $F_\mathrm{var}$ using Eq.~\ref{eq: covariance weighting} for 431 sources (d.o.f.~$=846$). The error bars show the $1\sigma$ dispersion from bootstrapping.}
\label{fig: vsh_offsets_weighting_effect_k}
\end{figure}

\begin{figure}
\includegraphics[width=\columnwidth]{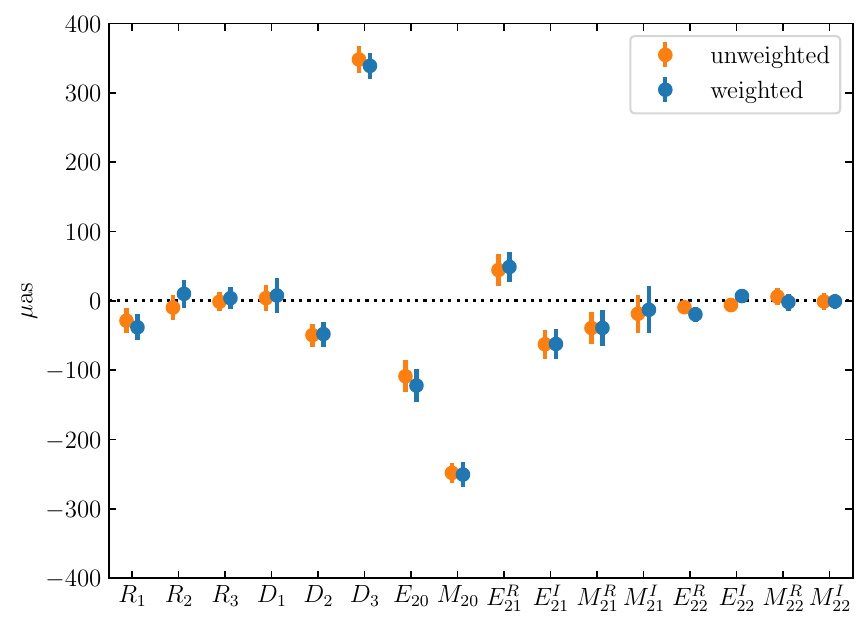}
\caption{Sixteen transformation parameters between ICRF3\,$X/Ka$ and {\it Gaia}~DR3 before and after weighting by $F_\mathrm{var}$ using Eq.~\ref{eq: covariance weighting} for 386 sources (d.o.f.~$=756$). The error bars show the $1\sigma$ dispersion from bootstrapping. The large distortions in $X/Ka$ are a known issue for ICRF3 \citep[compare with Fig.~19 in][]{2020A&A...644A.159C}.}
\label{fig: vsh_offsets_weighting_effect_xka}
\end{figure}

Repeating the bootstrap-based stability analysis on all 1938 objects, we found that weighting the covariance matrix using Eq.~\ref{eq: covariance weighting} resulted in a substantial improvement in frame distortions and stability, comparing $S/X$ to {\it Gaia} (Fig.~\ref{fig: vsh_offsets_weighting_effect_sx}), with an average improvement of a factor of 2 in the distortions and instabilities (Table~\ref{tab: rdem sx weighted}). The improvement is less notable in $K$, although the distortions and instabilities are slightly reduced for most VSH coefficients (Fig.~\ref{fig: vsh_offsets_weighting_effect_k} and Table~\ref{tab: rdem k weighted}), which is consistent with the weaker dependence of the covariance scaling on $F_\mathrm{var}$ because objects selected in $K$ are more likely to be blazars. ICRF3\,$X/Ka$ continues to be dominated by the distortions noted by \citet{2020A&A...644A.159C}, so there is no significant advantage to scaling the covariance matrices (Fig.~\ref{fig: vsh_offsets_weighting_effect_xka} and Table~\ref{tab: rdem xka weighted}).

\begin{table}
\caption{Frame distortions and instabilities in $\mu$as between ICRF3\,$S/X$ and {\it Gaia}~DR3 before and after weighting for 1938 sources (d.o.f.~$=3860$).}\label{tab: rdem sx weighted}
\centering
\begin{tabular}{crrrr}
\hline
\hline \\ [-0.36cm]
Parameter & $\hat{\mu}$ & $\hat{\mu}_w$ & $\hat{\sigma}$ & $\hat{\sigma}_w$ \\
\hline \\[-0.33cm]
$R_1$ & $+9$ & $+20$ & $26$ & $15$ \\ [0.05cm]
$R_2$ & $+3$ & $-9$ & $34$ & $15$ \\ [0.05cm]
$R_3$ & $-44$ & $+8$ & $42$ & $14$ \\ [0.05cm]
$D_1$ & $+24$ & $+22$ & $36$ & $21$ \\ [0.05cm]
$D_2$ & $-48$ & $-19$ & $36$ & $17$ \\ [0.05cm]
$D_3$ & $-36$ & $-25$ & $29$ & $15$ \\ [0.05cm]
$E_{20}$ & $-32$ & $-46$ & $34$ & $18$ \\ [0.05cm]
$M_{20}$ & $-60$ & $-14$ & $32$ & $16$ \\ [0.05cm]
$E^\mathrm{Re}_{21}$ & $-50$ & $-28$ & $34$ & $20$ \\ [0.05cm]
$E^\mathrm{Im}_{21}$ & $+41$ & $-12$ & $35$ & $18$ \\ [0.05cm]
$M^\mathrm{Re}_{21}$ & $+14$ & $-1$ & $50$ & $20$ \\ [0.05cm]
$M^\mathrm{Im}_{21}$ & $-89$ & $-4$ & $49$ & $31$ \\ [0.05cm]
$E^\mathrm{Re}_{22}$ & $-42$ & $-33$ & $32$ & $10$ \\ [0.05cm]
$E^\mathrm{Im}_{22}$ & $-11$ & $+8$ & $15$ & $9$ \\ [0.05cm]
$M^\mathrm{Re}_{22}$ & $-7$ & $+6$ & $16$ & $10$ \\ [0.05cm]
$M^\mathrm{Im}_{22}$ & $-3$ & $+1$ & $21$ & $10$ \\ [0.05cm]
\hline \\ [-0.33cm]
Average$^\dagger$ & $32$ & $16$ & $33$ & $16$ \\
\hline
\hline \\ [-0.2cm]
\end{tabular}
\tablefoot{Distortions are defined as the mean value of the bootstrap samples, and the standard deviation (dispersion) of the samples quantifies the instabilities. $^\dagger$The average of the distortions is taken over the absolute values.}
\end{table}

\begin{table}
\caption{Frame distortions and instabilities in $\mu$as between ICRF3\,$K$ and {\it Gaia}~DR3 before and after weighting for 431 sources (d.o.f.~$=846$).}\label{tab: rdem k weighted}
\centering
\begin{tabular}{crrrr}
\hline
\hline \\ [-0.36cm]
Parameter & $\hat{\mu}$ & $\hat{\mu}_w$ & $\hat{\sigma}$ & $\hat{\sigma}_w$ \\
\hline \\[-0.33cm]
$R_1$ & $+25$ & $+6$ & $23$ & $21$ \\ [0.05cm]
$R_2$ & $-13$ & $-12$ & $26$ & $20$ \\ [0.05cm]
$R_3$ & $+2$ & $+17$ & $19$ & $19$ \\ [0.05cm]
$D_1$ & $+59$ & $+25$ & $35$ & $32$ \\ [0.05cm]
$D_2$ & $-50$ & $-59$ & $21$ & $22$ \\ [0.05cm]
$D_3$ & $-27$ & $-32$ & $23$ & $18$ \\ [0.05cm]
$E_{20}$ & $-63$ & $-60$ & $28$ & $23$ \\ [0.05cm]
$M_{20}$ & $-16$ & $+3$ & $36$ & $28$ \\ [0.05cm]
$E^\mathrm{Re}_{21}$ & $+45$ & $+17$ & $33$ & $25$ \\ [0.05cm]
$E^\mathrm{Im}_{21}$ & $+49$ & $+48$ & $35$ & $29$ \\ [0.05cm]
$M^\mathrm{Re}_{21}$ & $-69$ & $-56$ & $27$ & $28$ \\ [0.05cm]
$M^\mathrm{Im}_{21}$ & $+7$ & $-12$ & $44$ & $42$ \\ [0.05cm]
$E^\mathrm{Re}_{22}$ & $-11$ & $-12$ & $17$ & $14$ \\ [0.05cm]
$E^\mathrm{Im}_{22}$ & $+15$ & $+20$ & $11$ & $11$ \\ [0.05cm]
$M^\mathrm{Re}_{22}$ & $-19$ & $-24$ & $15$ & $14$ \\ [0.05cm]
$M^\mathrm{Im}_{22}$ & $+39$ & $+29$ & $17$ & $13$ \\ [0.05cm]
\hline \\ [-0.33cm]
Average$^\dagger$ & $32$ & $27$ & $26$ & $22$ \\
\hline
\hline \\ [-0.2cm]
\end{tabular}
\tablefoot{Distortions are defined as the mean value of the bootstrap samples, and the standard deviation (dispersion) of the samples quantifies the instabilities. $^\dagger$The average of the distortions is taken over the absolute values.}
\end{table}

\begin{table}
\caption{Frame distortions and instabilities in $\mu$as between ICRF3\,$X/Ka$ and {\it Gaia}~DR3 before and after weighting for 386 sources (d.o.f.~$=756$).}\label{tab: rdem xka weighted}
\centering
\begin{tabular}{crrrr}
\hline
\hline \\ [-0.36cm]
Parameter & $\hat{\mu}$ & $\hat{\mu}_w$ & $\hat{\sigma}$ & $\hat{\sigma}_w$ \\
\hline \\[-0.33cm]
$R_1$ & $-29$ & $-38$ & $18$ & $19$ \\ [0.05cm]
$R_2$ & $-10$ & $+10$ & $19$ & $20$ \\ [0.05cm]
$R_3$ & $-1$ & $+4$ & $14$ & $16$ \\ [0.05cm]
$D_1$ & $+4$ & $+8$ & $19$ & $26$ \\ [0.05cm]
$D_2$ & $-49$ & $-48$ & $17$ & $18$ \\ [0.05cm]
$D_3$ & $+348$ & $+339$ & $19$ & $19$ \\ [0.05cm]
$E_{20}$ & $-109$ & $-122$ & $23$ & $24$ \\ [0.05cm]
$M_{20}$ & $-248$ & $-251$ & $14$ & $18$ \\ [0.05cm]
$E^\mathrm{Re}_{21}$ & $+44$ & $+49$ & $23$ & $22$ \\ [0.05cm]
$E^\mathrm{Im}_{21}$ & $-63$ & $-62$ & $21$ & $22$ \\ [0.05cm]
$M^\mathrm{Re}_{21}$ & $-39$ & $-39$ & $23$ & $25$ \\ [0.05cm]
$M^\mathrm{Im}_{21}$ & $-19$ & $-13$ & $28$ & $34$ \\ [0.05cm]
$E^\mathrm{Re}_{22}$ & $-9$ & $-20$ & $9$ & $10$ \\ [0.05cm]
$E^\mathrm{Im}_{22}$ & $-6$ & $+7$ & $10$ & $10$ \\ [0.05cm]
$M^\mathrm{Re}_{22}$ & $+6$ & $-2$ & $12$ & $12$ \\ [0.05cm]
$M^\mathrm{Im}_{22}$ & $-1$ & $-1$ & $12$ & $11$ \\ [0.05cm]
\hline \\ [-0.33cm]
Average$^\dagger$ & $62$ & $63$ & $17$ & $19$ \\
\hline
\hline \\ [-0.2cm]
\end{tabular}
\tablefoot{Distortions are defined as the mean value of the bootstrap samples, and the standard deviation (dispersion) of the samples quantifies the instabilities. $^\dagger$The average of the distortions is taken over the absolute values.}
\end{table}

We note that scaling the astrometric covariance might be interpreted as suggesting a variability-dependent, per-source position uncertainty correction factor, but we caution against this interpretation. The modification of the individual source position errors in a way that effectively scales the VSH fit covariance leads to covariance-normalized individual position offsets $X$ that are heavily skewed to low values because many sources with a low fractional variability are nonetheless blazars or sources that otherwise do not have notable position offsets. Any per-source position error correction method should ideally produce values of $X$ following a Rayleigh distribution with $\sigma=1$. The covariance scaling method shown above only accounts for the average behavior of sources as a class. This allows optimal weighting of the ties between celestial reference frames.

\section{Discussion} \label{sec: discussion}
We reiterate that the bootstrap method we employed is statistically equivalent to sampling the astrometric behavior of a single set of objects across multiple epochs after accounting for the statistical error. It is therefore a direct measure of the frame stability over time. Our results strongly support a picture in which the small line of sight to blazars improves their astrometric properties, and we have presented a method of optimally weighting all sources to improve frame distortions and stability, especially the stability of the $S/X$ frame. 

Nonetheless, we anticipate two objections to the prospect of creating a blazar-based celestial reference frame. The first objection is that as their photometric variability implies, blazars are dynamic objects that are expected to exhibit phenomena such as superluminal motion or changes in jet orientation over human timescales, a quality that might make a researcher hesitate to give blazars preference in reference frame work. The second objection is that blazars are relatively rare objects, and there might therefore be concern about their availability and the constraints placed by this lack of availability on the creation of a future reference frame. In Sect.~\ref{subsec: longterm} we review the dynamical processes observed in jets and discuss why blazars appear to be astrometrically stable in spite of these effects; and in Sect.~\ref{subsec: availability} we comment on the availability of blazar-like sources for the creation of a celestial reference frame.

\subsection{Long-term astrometric stability of blazars} \label{subsec: longterm}
It is important to note that changes in the line-of-sight jet angle of radio AGN have been observed on human timescales. Using 20 years of VLBI observations, \citet{2023Natur.621..711C} found that the jet in M87 precesses around an angle of $\sim1\fdg25$ over the span of $\sim11$~yr. \citet{2023A&A...672L...5V} similarly reported precession with a period of $\sim7$~yr in M81, with an amplitude of $7^\circ$. \citet{2023A&A...672L...5V} also reported, however, a slow and apparently linear drift in the jet angle of $0\fdg5$~yr$^{-1}$, which they showed can be interpreted as the true precession of the jet with a long but almost unconstrained period of between 200 and 1800~yr, indicating that the 7~yr sinusoidal variation is nutation. While precession can change the alignment of jets by a few degrees on timescales of decades, jets can reorient on much longer timescales. \citet{2013ApJ...768...11B} presented a model in which the random orientation of a newly formed geometrically thin accretion disk induces a torque that causes the black hole spin axis to change on $\sim$~Myr timescales. X-ray data of clusters indicate that intercluster heating is approximately isotropic, suggesting that disparate episodes of thin-disk formation can swivel the black hole spin axis across $\sim4\pi$\,sr over $\mathcal{O}(100)$~Myr. 

With nutation on $\mathcal{O}(10)$\,yr timescales and precession on $\mathcal{O}(>100)$\,yr timescales, the Doppler factors $\delta$ of blazars should remain high over human timescales. Given a typical blazar with Doppler and Lorentz factors around $\delta \sim \Gamma \sim 10$ \citep[e.g.,][also \citealt{2019ApJ...874...43L}]{2017MNRAS.466.4625L, 2020ApJ...897...10Z}, corresponding, for example, to $\beta=0.995$ and line-of-sight angle $\theta\sim6$\textdegree, the Doppler factor is expected to vary between $\delta\sim4$ and $\delta\sim20$ for a projected nutation amplitude of 5\textdegree. The flux-beaming factor, which scales as $\delta^n$, is therefore expected to vary around a mean value of $\sim90$ within a factor of five, given $n=2$ \citep{2007ApJ...658..232C}, and the jet is expected to continue to dominate over the accretion disk at visual wavelengths, as indicated by the redder color of blazars \citep[e.g., Fig.~4 in][]{2022ApJ...939L..32S}. The optical-radio offset is then the projected distance between the optical photocenter along the jet and the radio peak that defines the VLBI position, ostensibly the frequency-dependent position at which the core becomes optically thin. In sources with estimated radio and optical positions along the jet, the optical position is downstream of the radio position about 80\% of the time at distances of $\mathcal{O}(10)$\,pc \citep[e.g.,][]{2019MNRAS.485.1822P}. Nutation with decade-scale periods is therefore unlikely to change the optical positions of jets, which are at comparable light-crossing time distances, so nutation is expected to mainly affect the apparent positions of the radio cores. The recent review of the well-studied jet in M87 by \citet{2024A&ARv..32....5H} summarized the deprojected core positions of the jet as a function of frequency (their Fig.~8), showing that the frequency-dependent core is within $\sim0.1$~pc of the black hole. Then, for the typical blazar described above, the optical-radio offset is expected to be about 0.1~mas for a $z=1$ source ($h=0.7$) and will remain stable to within about 1~$\mu$as over the nutation amplitude.

It is likely, however, that the deprojected radio core positions of typical $z\sim1$ sources lie at considerably larger distances from the central engine than what is observed in nearby M87, because VLBI observations are sensitive to far more of the extended jet. The preceding discussion demonstrated only that nutation, which occurs over human timescales but is limited to jet structures within the light-crossing time, is likely to be a negligible source of astrometric variation in blazars. Astrometric variation in the jet might therefore be limited to superluminal motion or to pseudo-motion induced by flares in quasi-static shocks.

In the case of superluminal motion, the maximum apparent velocity is $\beta_\mathrm{app.}\sim\Gamma=10$ and corresponds to $\theta\sim6$\textdegree, the jet line-of-sight angle in our example. If the structures corresponding to the VLBI and optical positions are superluminal to this degree, then they are expected to exhibit proper motions up to $\sim0.4$~mas~yr$^{-1}$, comparable to the average total {\it Gaia} proper motion uncertainty of our sample. The proper motion uncertainties in about 20\% of the sample are small enough for such large superluminal motions to be detectable, but as outlined in Sect.~\ref{subsec: sample}, \citet{2022ApJ...939L..32S} removed objects with formally significant proper motions because these objects were shown by \citet{2022A&A...660A..16S} and \citet{2022ApJ...933...28M} to often be multisources such as dual AGN. We therefore restored these objects to the sample, which also entailed allowing those with {\it Gaia} counterparts not in {\it Gaia}-CRF3, yielding 16 objects in addition to the original 1938. We calculated the covariance-normalized proper motions $\chi$ using Eq.~1 in \citet{2022ApJ...933...28M} and determined the statistical consistency of $\chi$ between blazars and nonblazars using the Anderson-Darling test. When nonblazars are defined as objects with $F_\mathrm{var} < 0.2$ and blazars as those with $F_\mathrm{var} > 0.4$, there is no compelling evidence that the distributions of total proper motions $|\bm{\mu}|$ are different ($p=0.08$), but mild evidence that the covariance-normalized proper motions $\chi$ do differ ($p=0.03$). When we relaxed the definition of a blazar to objects with $F_\mathrm{var} > 0.2$, the statistical power improved and yielded a highly significant ($p\sim0$) difference in total proper motions, with blazars having a mean $\left<|\bm{\mu}|\right> \sim 0.4$~mas~yr$^{-1}$ and nonblazars having $\left<|\bm{\mu}|\right> \sim 0.3$~mas~yr$^{-1}$, but the difference in covariance-normalized proper motion is insignificant ($p=0.24$), which indicates that the apparently higher proper motions observed in the blazar class are an artifact of larger measurement errors. The sum of the average proper motion covariance matrix in the nonblazars is 0.23~mas$^2$~yr$^{-2}$, and for blazar-like objects with $F_\mathrm{var} > 0.2$ the sum is 0.31~mas$^2$~yr$^{-2}$, although it is 0.23~mas$^2$~yr$^{-2}$ when blazars are defined as $F_\mathrm{var} > 0.4$. This explains the difference in the significance tests. When we instead compare the mean proper motions of objects with significant values of $\chi$, we find that for $\chi > 3, 4, 5$ the mean proper total motion of nonblazars is $0.81\pm0.18$, $0.58\pm0.16$, $1.11\pm0.41$ mas~yr$^{-1}$ and for blazars ($F_\mathrm{var} > 0.2$), it is  $0.63\pm0.12$, $0.69\pm0.28$, $0.46\pm0.22$ mas~yr$^{-1}$. There is therefore no significant evidence that blazars exhibit systematically larger proper motions than nonblazars for objects with (formally) significant proper motions.

The question remains, however, why {\it Gaia} does not appear to be sensitive to superluminal motions in the jets of these objects. The likely explanation is that given the moderate redshifts of ICRF objects, {\it Gaia} measures the full jet, which dominates over individual superluminal knots. For example, the jet in M87 has several superluminal knots \citep[e.g.,][]{1999ApJ...520..621B}, but they are all much fainter than the full jet, which would be subsumed within the {\it Gaia} point spread function at $z = 1$. The optical astrometric stability of blazars might therefore not necessarily extend to low redshift where the centroid may correspond to a superluminal component of the jet. This is already true of low redshift sources in general, however, which can induce spurious proper motions in {\it Gaia} astrometry due to source extent \citep{2022A&A...660A..16S, 2022ApJ...933...28M}. Fortunately, it is generally straightforward to remove extended sources, in the case of {\it Gaia} by excluding sources with large \texttt{phot\_bp\_rp\_excess\_factor}.

While we did not detect significant differences in the apparent proper motions of blazar- and quasar-like ICRF sources, stochastic flares might cause variability-induced motion \citep[e.g.,][]{2016ApJS..224...19M}, which in {\it Gaia} data manifest as a significant \texttt{astrometric\_excess\_noise} parameter. In the original sample of 1938 objects, we again found no significant differences between blazars and quasars. Objects with $F_\mathrm{var} < 0.2$,  $F_\mathrm{var} > 0.2$, and $F_\mathrm{var} > 0.4$ all have a mean \texttt{astrometric\_excess\_noise} of $\sim0.2$~mas and a mean \texttt{astrometric\_excess\_noise\_sig} of $\sim0.3$, which is close to the expected value of $\sqrt{1/2\pi}\sim0.4$ for a normal distribution with negative values set to zero.\footnote{\url{https://gea.esac.esa.int/archive/documentation/GDR3/Gaia_archive/chap_datamodel/sec_dm_main_source_catalogue/ssec_dm_Gaia_source.html}} About $50\%$ of the objects in both classes have \texttt{astrometric\_excess\_noise}~$=0$, and an Anderson-Darling test did not indicate inconsistencies in the astrometric excess noise or its significance between the two classes. In VLBI data, flares might induce strong changes in the apparent position angle of blazar jets, which are common and might arise because the jet opening angle is comparable to the viewing angle \citep[e.g.,][]{2013AJ....146..120L, 2025MNRAS.537..978K}. For a typical position angle standard deviation of $\sim10^\circ$ and core offset of $\sim1$~mas \citep{2013AJ....146..120L}, we might expect a VLBI position dispersive component of $\sim0.2$~mas. However, \citet{2024A&A...684A..93L} showed that ICRF objects with the lowest VLBI position variability exhibit both the highest fractional photometric variability and the highest prevalence of $\gamma$-ray detection. This indicates that blazars are in fact more astrometrically stable in VLBI data, so any astrometric variability caused by large changes in jet position angle does not appear to be significant.

It is clear from the considerations discussed above that although AGN jets are highly dynamic phenomena that exhibit precession, nutation, superluminal motion, and stochastic flaring, these phenomena likely do not significantly detract from the astrometric stability gained by giving preference to sources with jets aligned close to the line of sight where relativistic Doppler beaming enhances the consistency between the optical and radio positions and any inconsistencies are minimized in projection. These arguments explain the empirical fact, demonstrated in the preceding sections, that selection on blazars yields a dramatic improvement in the stability of the celestial reference frame, an improvement that exceeds the stability of the current set of ICRF defining sources, as well as the finding that weighting by source variability allows for use of all sources that have variability information. This confirms the prediction made by \citet{2022ApJ...939L..32S}.

\subsection{Availability of sources} \label{subsec: availability}
As noted in Sect.~\ref{subsec: sample}, the sample we used is only a subset of the sources available to construct a frame that takes source variability into account. While we focused on a sample of 180 likely blazars from ICRF3 identified by \citet{2022ApJ...939L..32S}, incorporation of other multi-epoch photometry and larger parent samples of VLBI sources will significantly increase the number of identified blazars. Moreover, as we showed in Sect.~\ref{subsec: weighting}, knowledge of the photometric variability of sources allows for a simple and effective means to weight the frame alignment, and this flexibility can create a sample of defining sources that is more uniformly distributed across the sky. This is a priority of the current ICRF. With the upcoming {\it Gaia} DR4 (2026) and {\it Gaia} DR5 (2031) releases, which will provide two and four times as many months of data as DR3, respectively, it is likely that the completeness of blazar catalogs based on {\it Gaia} data will improve considerably. Several other ground-based surveys such as ZTF and LSST will further improve completeness, and it is likely that infrared data from the WISE/NEOWISE catalogs, which span over ten years, will be useful as well. Finally, dedicated near-infrared monitoring campaigns of VLBI sources within the Galactic plane, where reddening prevented detection by {\it Gaia}, might prove invaluable for the source selection or weighting of a future infrared astrometric mission such as {\it Gaia}NIR \citep{2021ExA....51..783H}.

\section{Conclusions} \label{sec: conclusions}
Using ICRF3 sources with measured fractional photometric variability $F_\mathrm{var}$ in {\it Gaia}~DR3, we have conducted the first study of the effect of astrophysical source type (quasar-like, intermediate, or blazar-like) on the stability of the celestial reference frame. Using the 16-parameter vector spherical harmonic model employed by \citet{2020A&A...644A.159C} to characterize ICRF3, we quantified the time-dependent stability of the reference frame by noting that random sampling with replacement (the bootstrap) is statistically equivalent to sampling a particular class of source (e.g., blazars) over multiple epochs, given the independence between sources. Our main conclusions are:

\begin{enumerate}
\item Photometric variability strongly predicts stability of the celestial reference frame, with the most variable, blazar-like objects defining a frame that is six times more stable than the most photometrically constant, quasar-like objects, comparing ICRF3\,$S/X$ to {\it Gaia}. In the higher-frequency $K$ and $X/Ka$ bands, this difference persists, but is smaller, likely because the sources are a priori more likely to be blazars. 
\item At $S/X$, blazar-like objects define a frame that is over twice as stable as even the ICRF3 defining sources. This suggests that selecting defining sources based on photometric variability is superior to a selection based on source compactness or apparent VLBI positional stability, as was the case for ICRF3 and previous ICRFs, although we emphasize that we showed that blazar-like objects have more stable VLBI positions in \citet{2024A&A...684A..93L}. A mild improvement of a factor of 1.4 is also seen in $X/Ka$, although blazars perform indistinguishably from the defining sources in $K$.
\item We showed that scaling the original VLBI source position covariances by $F_\mathrm{var}^{-x}$ provides a remarkably simple means of optimally weighting the tie between celestial reference frames. The value of the exponent $x$ is steepest for $S/X$ and flattest for $X/Ka$, as expected if blazars are more prevalent among sources selected at $K$ and $Ka$. The values we determined in Sect.~\ref{subsec: weighting} leave no unexplained variance in the 16-parameter VSH transformations between frames and substantially improve the distortions and stability of the link between ICRF3\,$S/X$ and {\it Gaia}.
\end{enumerate}

Our results build upon the discovery by \citet{2022ApJ...939L..32S} that blazars are systematically less likely to exhibit optical-radio position offsets, and on the discovery by \citet{2024A&A...684A..93L} that blazars also exhibit significantly less astrometric variability in multi-epoch VLBI data \citep[see also Sect.~6 in][]{2024A&A...691A..35B}. In addition to quantifying the effect of blazar source selection and weighting on frame stability, we further explored the physical mechanisms of this stability, which superficially seems counterintuitive given the dynamic nature of blazars. We argued that time-varying phenomena such as jet precession and nutation are unlikely to be important on human timescales, and that the effect of superluminal motion is likely negligible for moderate-redshift objects, which comprise the majority of the ICRF. These arguments provide a theoretical explanation for the empirical fact, demonstrated here and predicted by \citet{2022ApJ...939L..32S}, that blazars are inherently more astrometrically stable, and that source variability can be used to optimally tie celestial reference frames.

As discussed by \citet{2024A&A...684A..93L} and emphasized by \citet{2025A&A...695A.135L}, photometric variability monitoring campaigns are much less observationally expensive than VLBI monitoring campaigns, so regular observations by networks of one-meter-class telescopes can therefore serve to augment reference frame work by better distinguishing the astrometrically stable blazars from the unstable quasars. With nearly 22,000 known VLBI sources \citep{2025ApJS..276...38P}, our results show that determining the photometric variability properties of these sources has the potential to significantly improve the precision and stability of the celestial reference frame.

\begin{acknowledgements}
The authors thank the anonymous referee for their thoughtful review that significantly improved the quality of this work. N.J.S.\ gratefully acknowledges V.~Makarov for helpful mathematical discussions and clarifications. This work has made use of data from the European Space Agency (ESA) mission {\it Gaia} (\url{https://www.cosmos.esa.int/Gaia}), processed by the {\it Gaia} Data Processing and Analysis Consortium (DPAC, \url{https://www.cosmos.esa.int/web/Gaia/dpac/consortium}). Funding for the DPAC has been provided by national institutions, in particular the institutions participating in the {\it Gaia} Multilateral Agreement. 

This work made use of Astropy:\footnote{http://www.astropy.org} a community-developed core Python package and an ecosystem of tools and resources for astronomy \citep{astropy:2013, astropy:2018, astropy:2022}, as well as TOPCAT \citep{2005ASPC..347...29T}.

\end{acknowledgements}

\bibliographystyle{aa}
\bibliography{aa56690-25}

\end{document}